# More religion means less science:
## An International comparison of the relations between religious beliefs and levels of and attitudes to scientific knowledge

Yves Gingras[1] and Kristoff Talin[2]

[1] *gingras.yves@uqam.ca*
Centre interuniversitaire de recherche sur la science et la technologie (CIRST), Université du Québec À Montréal, C.P. 8888, Suc Centre-Ville, Montréal, Qc, Canada, H3C 3P8.

*2 christophe.talin@univ-lille.fr*

*CNRS-Clersé, MESHS2, rue des Canonniers - 59000 Lille, France*

## Abstract

This research presents the results of a comparative analysis of the links between religious practices and beliefs and levels of scientific knowledge. Based on secondary analyses of survey data in the European Union (Eurobarometers 2005 and 2010) and the United States (Pew Research Center 2018), we show that, regardless of the country, correlations suggest that the more individuals identify with a religion and the more intensely they practice that religion, the less scientifically literate they are, as measured in standard tests. Moreover, scientific representations are also related to an individual's religious outlook. The more individuals adhere to a religion, the less they have positive attitudes towards science. The conclusion suggests possible interpretations of theses correlations.

**Introduction**

Since the early 1980s, there has been a new trend promoting "dialogue" between science and religion, claiming that conflicts between these two modes of thought have been greatly exaggerated. These ideas have stimulated new research to see whether or not scientists perceive any conflict between science and religion. For instance, Ecklund and Park [1] have conducted interviews with university scientists. Their results show that most respondents do not perceive any conflict between science and religion. When scientists are more integrated, through their practices, with their religion, they are even less likely to assert the existence of such a conflict. According to the authors, these results contradict earlier findings and show that religion and science are not incompatible. They extended their analysis to eight countries (France, Hong Kong, India, Italy, Taiwan, Turkey, United Kingdom, United States) thus providing a comparative perspective [2]. Based on a survey of physicists and biologists the authors reaffirm their previous results and conclude that most scientists consider religion and science as operating in separate spheres.

Ecklund's team [2] also show that scientists are much less religious – in terms of practices and beliefs – than ordinary citizens. This is, of course, consistent with the many analysis conducted since the beginning of the 20$^{th}$ century. In 1914, for example, the American psychologist James H. Leuba conducted a survey [3] in which he showed that only 27% of the American scientific elite at that time believed in the existence of a personal God. He updated his survey in 1933 and showed that this percentage had fallen to 15% [4]. Fifty years later, a new survey, based on Leuba's method for purposes of comparison, was conducted among American scientists [5]. The results confirmed the decline: from 15% in 1933, the proportion of believers in a personal God had declined to 7%. The belief rate of American scientists in a personal God thus fell by 50% once, between 1914 and 1933, and then by half again between 1933 and 1998. More generally, the religious sphere is still characterized by a marked decline in traditional forms of practice (belonging to a religion, attending religious services, prayer [6]) and the emergence of new forms of religious expression (belief in para-sciences, the development of "emotional forms" [7]). In a recent study, Inglehart [8] observed that 43 out of 49 countries he studied saw a decline in religion between 2007 and 2019, the most dramatic shift noted being among the American public.

Research on perceptions of conflict conducted among American undergraduate students by Scheitle [9] shows results quite similar to those of Ecklund [2]. The majority of students do not see the religious and scientific spheres as being in conflict. There are, however, interesting variations according to the level of religious integration: on one hand, highly integrated students – as well as those who are the most religiously conservative – are more likely to see a conflict between religion and science; on the other hand, at equal levels of religious integration, students in a religious institution are less likely to see a conflict between science and religion than students in a secular school. These results suggest that enrollment in a religious institution may help reduce the perception of tensions between religion and science.

In Belgium, a survey conducted in 2011, among a representative sample of high school students in their final year, also gives similar results [10]. The interest of this research is that it highlights different types of conceptions of science that are strongly correlated with what the author calls the "register of students' convictions". It is possible to position these different types on a continuum where at one end we have a high degree of adherence to religious values – as well as a rejection of rationalist criticism – associated with a very weak recognition of the autonomy and specificity of science. At the other end of the spectrum, we find agnostic or atheist students who see science as very autonomous from religion.

Finally, Evans [11] shows that, in the United States, levels of religious integration and opinions about science are correlated. The most integrated members of the different Protestant movements, like conservative Catholics, side in favor of religion if a conflict between science and religion arises over their world view. These two groups also try to limit the influence of scientists on moral issues in the public sphere. In a subsequent article, Evans [12] shows that these conclusions are reinforced when the respondent belongs to a religious group that promote a fundamentalist reading of the Bible.

Most of these studies inquired about how scientists, or college and university students, *perceive* the possible relations between science and religion. One limitation of these studies based only on representations is that given the distinction between facts and norms, and between reality and representation of reality, they do not prove that conflicts did not or could not in fact arise in certain social contexts. Another limitation is that the philosophical question of the epistemic difference between science and religion cannot be reduced to that of knowing what scientists think the relations between these two domains are or should be. For example, it is well known that specific conflicts between specific sciences (astronomy, geology, biology) and specific religions (Christianism, Islam) have existed at different times and places since the 17$^{th}$ century despite discourses alleging that 'well understood' religions cannot 'really' conflict with 'true' science [13].

A different approach to the study of the relations between science and religion, that goes beyond the analyses of representations or perceptions of these relations, consists in finding independent measures of religious affiliations and practices on the one hand and of level of scientific knowledge and attitudes toward science on the other and then looking at the possible correlations between them.

Starting from the hypothesis that beliefs (scientific or religious) have cognitive effects – otherwise one might wonder whether they are really sincere and not simply superficial beliefs – we wanted to see if, at the macro-social level of the general population of different countries, knowledge of and attitudes towards science vary according to the type and level of religious belief in addition to the usual sociodemographic variables. For even if someone may personally think that there is or should be no conflicts between science and religion, it may actually be the case that, there could indeed exist, at the macro-social level, negative correlations between intense beliefs about religion and the level of knowledge of and attitudes toward science.

**Data sources**

In order to better understand the ways in which religious beliefs and scientific knowledge are articulated in contemporary Western societies, we have conducted a secondary analysis of existing European and American surveys, whose wealth of data has been under-exploited. These surveys are:

- the Eurobarometer surveys conducted in 2005 and 2010 in European Union member countries. That is, the 25 countries in 2005 plus Romania and Bulgaria (members in 2007) and Croatia (member in 2013). The first, No. 63.1, dates from 2005 and covers Europeans, science and technology [14]. The second, No. 73.1, dates from 2010 and focuses on biotechnology [15]. The data include variables on religious adherence, levels of religious practice and belief in the existence of a God. This makes it possible to take account of different dimensions of the relationship to religion. The knowledge and representation of science are approached through different questions on basic scientific facts and scales of attitudes toward science;
- the survey conducted by the Pew Research Center in 2018 in the United States [16]. This data set tells us about religious affiliation, the intensity of religious practice, the frequency of prayer, and the importance of religion in daily life. Scientific knowledge is measured using a series of nine variables to construct a scientific knowledge score.

In all these surveys, the religious identity is self-declared by the respondents. Thus, a "Catholic", a "Protestant" an "Evangelic" or "Born again Christian" etc. is a person who declared that particular identity. Likewise, the intensity of adhesion to these different religions is self-declared by responding to questions on the frequency of practices like going (regularly or not) to a church, following rites, etc.

In the following section, we first present the results of the analysis of the links between religious practices and beliefs and scientific knowledge. Section 2 deals with the influence of socio-demographic variables on these relationships. Section 3 discusses the links between representations of science and of its social effects (positive or negative) and religious practices and beliefs. In the conclusion, we briefly recall the main findings and limits of these surveys and suggest possible explanations of the correlations observed.

## 1. Links between practices, religious beliefs and scientific knowledge in Europe and the United States

This section presents the results of the analysis of the links between religious practices and beliefs and scientific knowledge based on the European (section 1.1) and American (section 1.2) surveys. In both cases, we observe negative correlations between the level of scientific knowledge and the intensity of religious practices and beliefs.

### 1.1 Europe

The level of scientific knowledge was measured by a series of 13 factual questions. Each respondent was given a score between 0 and 13 depending on the number of correct answers. Figure 1 shows the distribution of correct answers for all respondents to the 2005 Eurobarometer questionnaire. The average is 8.24, or 63%, with an asymmetric Gaussian distribution to the right. Depending on the point of view taken, one could emphasize the relatively good knowledge of basic scientific facts or, on the contrary, insist that a quarter of the sample had less than 50% of correct answers.

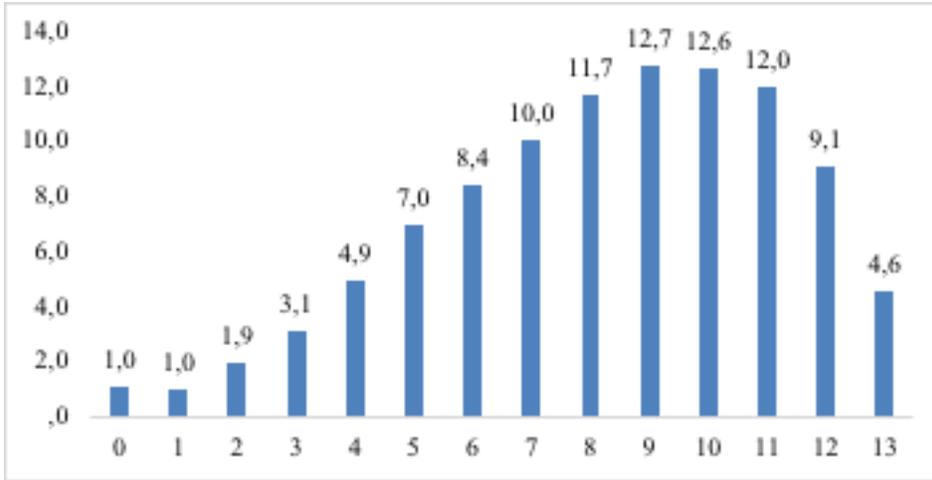

**Figure 1**
**Number of correct answers on the scientific knowledge scale**
**(N=27909, in percent)**

But what interests us here is less the level of knowledge *per se*, than its possible relation with other variables, particularly those related to religious practices and beliefs. It has long been known that the level of knowledge is strongly associated with the level of education. Not surprisingly, therefore, the average number of correct answers is lower for those with only a high school education than for those with a university degree. But what about the relationship with religious affiliation? Table 1 shows that there are significant variations in the level of scientific knowledge according to the declared religion. Those with no religion have the highest average, followed by Protestants and other religions (mostly Asian). All three groups score higher than the average of the whole sample. On the other hand, and in descending order, Catholics, Orthodox and Muslims score below average. Religion and level of scientific knowledge are therefore strongly correlated (Unless otherwise indicated, F-test are significant at the 0.000 level for all the tables included in this paper).

**Table 1**
**Average score on the scientific knowledge scale according to declared religious affiliation**

|  | Mean | Number |
|---|---|---|
| **Overall sample** | **8,24** | **27384** |
| No religion | 9,56 | 4653 |
| Protestant | 9,06 | 3958 |
| Other religions | 8,82 | 1705 |
| Catholic | 7,91 | 13076 |
| Orthodox | 6,80 | 3694 |
| Muslim | 6,00 | 298 |

In addition to religious affiliation, survey data allow us to measure intensity of practice, which is a relevant indicator of religious integration. One might, for example, declare oneself Catholic and never go to Church. Such an individual is clearly different from a declared Catholic who attends Church weekly. As Table 2 shows, the less Europeans practice any religion, the higher their scientific knowledge score, with a difference of almost three standard deviations between non-practitioners and weekly practitioners.

**Table 2**
**Average score on the scale of scientific knowledge according to the intensity of religious practice**

|  | Mean | Number |
|---|---|---|
| **Complete sample** | **8,25** | **27661** |
| At least once a week | 7,18 | 5271 |
| About once a month | 7,85 | 2458 |
| Two to three times a year | 8,18 | 7165 |
| Less often | 8,67 | 6016 |
| Never | 8,91 | 6751 |

Table 3 that joins religion and practices versus no religion, confirm the same trend.

**Tableau 3**
**Average score on the scale of scientific knowledge according to religion and level of practice**

|  | Mean | Number |
|---|---|---|
| **Complete sample** | **8,13** | **24301** |
| Religious, practicing | 7,87 | 18261 |
| Religious, non-practicing | 8,16 | 2652 |
| No religion | 9,50 | 3388 |

The practice indicator, which is a synthetic measure of the level of integration into a religion [17], is therefore negatively correlated with the level of scientific knowledge: the more individuals practice their religion, the lower their knowledge score. This variation is valid for all religions except for Muslims (Table 4). This result can probably be attributed to the small number of Muslims in the sample. Moreover, the results for Protestantism seem particularly significant. They are consistent with the thesis well known to sociologists of science, which associates the vigorous development of science in 17th century England with scholars belonging more often to Protestant sects [18]. Among religious individuals, Protestants have the highest level of scientific knowledge. Nevertheless, Protestants that practice the least still have a lower average than those declaring themselves to be without any religion.

**Table 4**
**Average score on the scale of scientific knowledge according to religion and level of practice**

|  | Mean | Number |
|---|---|---|
| **Complete sample** | **8,21** | **25566** |
| No Religion | 9,56 | 4653 |
| Less-practicing Protestant | 9,16 | 3245 |
| Monthly-practicing Protestant | 8,59 | 701 |
| Less-practicing Catholic | 8,25 | 7599 |
| Monthly-practicing Catholic | 7,43 | 5398 |
| Less-practicing Orthodox | 7,07 | 2497 |
| Monthly-practicing Orthodox | 6,24 | 1178 |
| Monthly-practicing Muslim | 6,75 | 61 |
| Less-practicing Muslim | 5,83 | 234 |

The indicator of belief in God measures a more subjective and less ritualistic dimension of religious feeling. One in two respondents believe in the existence of a God, one in three say there is a spiritual or living force, and 15% believe in neither (Table 5).

**Table 5**
**Percentage of subjective beliefs in a God (Vertical percentages)**

|  | Percentage | Respondents |
|---|---|---|
| I believe there is a God | 54 | 16628 |
| I believe that there is a spiritual or living force | 31 | 9508 |
| I don't believe in God or a spiritual force | 15 | 4435 |

Of course, subjective beliefs are strongly correlated with the religion to which one belongs (Table 6). However, minority groups – sometimes unexpected – do exist. For example, 10% of Protestants say they have no subjective belief in a God, while 39% of those without religion believe in a spiritual or living force. It is likely that in these cases, respondents were expressing a purely cultural conception of their religious identity.

**Table 6**
**Subjective beliefs in a God according to declared religion**
**(N=25092, horizontal percentages)**

|  | I believe there is a God | I believe that there is a spiritual or living force | I don't believe in God or a spiritual force |
|---|---|---|---|
| **Complete sample** | **56** | **29** | **15** |
| Catholic | 71 | 23 | 6 |
| Orthodox | 74 | 23 | 3 |
| Protestant | 48 | 42 | 10 |
| Muslim | 74 | 20 | 6 |
| No religion | 6 | 39 | 56 |

Cramer's V = 0,43

Table 7 shows that belief in a God or in a spiritual force, is correlated with level of scientific knowledge. For example, non-believing Protestants score highest among respondents who declare a religion. Interestingly, the belief in a God or spiritual force also affects the level of knowledge of persons without religion. Hence, those without a religion who believe in God score significantly lower (8.38) than those who declare themselves to be non-believers (9.69).

**Table 7**
**Average score on the scale of scientific knowledge according to the typology religion and belief**

|  | Mean | Number |
|---|---|---|
| **Complete sample** | **8,21** | **25092** |
| Catholic believing in God | 7,59 | 9093 |
| Catholic believing in a spiritual force | 8,77 | 2941 |
| Non-believing Catholic | 8,44 | 775 |
| Orthodox believing in God | 6,47 | 2679 |
| Orthodox believing in a spiritual force | 7,75 | 816 |
| Non-believing Orthodox | 7,54 | 126 |
| Protestant believing in God | 8,71 | 1866 |
| Protestant believing in a spiritual force | 9,41 | 1625 |
| Non-believing Protestant | 9,40 | 386 |
| Muslim believing in God | 6,16 | 205 |
| Muslim believing in a spiritual force | 6,36 | 56 |
| Muslim, non-believing in God | 5,44 | 16 |
| No religion, believing in God | 8,38 | 251 |
| No religion, believing in a spiritual force | 9,58 | 1742 |
| No religion, non-believing | 9,69 | 2515 |

On the basis of these European data, and as a partial conclusion, we can say that belonging to a religion, having a high level of religious practice and believing in God correlate with lower levels of scientific knowledge. Now let us look at the situation in USA.

### 1.2 United States

In the Pew Research U.S. survey, the level of scientific knowledge is gauged by a group of nine questions. For each of them, respondents were asked to choose the correct answer. On average, 52% of respondents answered correctly, but the dispersion of responses, depending on the question, is significant and follow a normal distribution. For example, while 75% were able to identify carbon dioxide as the gas produced as a result of combustion, only 37% identified nitrogen as the most common gas in the earth's atmosphere.

By counting all of the correct answers, a cumulative index can be constructed ranging from 0 (no correct answers) to 9 (all correct answers). 34% scored less than 4, while 26% scored between 4 and 5 and 40% scored 6 or higher (Figure 2).

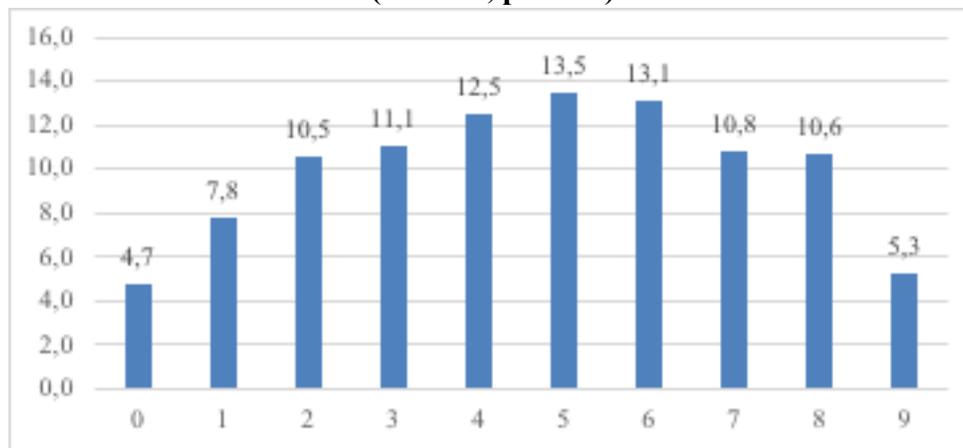

**Figure 2**
**Number of correct answers on the scientific knowledge scale**
**(N=2537, percent)**

As was the case in European Union countries, the level of scientific knowledge of Americans correlates with their religious practices and beliefs (Table 8, For the purposes of our analysis, the religions analyzed include only the five groups mentioned. Minority religions were excluded because of small numbers that do not allow for robust statistical analyses. In this table, religious integration is measured using an index composed of three variables (the importance of religion in life, the frequency of religious practice and the intensity of prayer). Individuals with high religious integration attend religious services at least once a week, pray daily and say that religion is very important in their lives. Those with low religious integration say that religion is of little or no importance in their lives, pray rarely or never, and rarely or never attend religious services. Other respondents are classified in the intermediate integration group. Thus, Catholics (3.89), Protestants (4.05) and those who report having no particular religion (4.21) are below the average knowledge score (4.42). They differ markedly from agnostics (6.24) and atheists (7.04) who have much higher scores. The group declaring no religion has an intermediate score. Agnostics, who are non-believers but less averse to religion than atheists, follow those who have no religion.

**Table 8**
**Average score for correct answers to knowledge questions based on different measures of the degree of identification with a religion**

| | |
|---|---|
| **Declared Religion (N=2264)** | |
| **Complete Sample** | **4,42** |
| Catholic | 3,89 |
| Protestant | 4,05 |
| None | 4,21 |
| Agnostic | **6,24** |
| Atheist | **7,04** |
| **Degree of Religious Practice** | |
| **Complete Sample** | **4,48** |
| At Least Once a Week | 4,09 |
| At Least Once a Month | 4,18 |
| Several Times a Year | 4,19 |
| Rarely | **4,49** |
| Never | **5,06** |
| **Prayer frequency** | |
| **Complete Sample** | **4,55** |
| At Least Once a Day | 3,93 |
| At least Once a Month | 4,29 |
| Rarely | **4,92** |
| Never | **5,94** |
| **Importance of Religion in Your Life** | |
| **Complete Sample** | **4,45** |
| Very Important | 3,77 |
| Quite Important | 4,08 |
| Not Very Important | **4,79** |
| Not Important at All | **5,67** |
| **Religious Integration** | |
| **Complete Sample** | **4,55** |
| Strong | 4,08 |
| Intermediate | 4,15 |
| Weak | **5,68** |

These results show that the more people are integrated into a religion, the less they perform in terms of scientific knowledge. Moreover, all indicators of the level of religious practice – prayer and the importance of religion in daily life – vary in the same direction: the more an individual practices, prays or is religious, the lower his or her scientific knowledge score.

The combination of two variables – the religion to which one belongs and the level of religious integration – offers significant contrasts (Table 9. The numbers for the statistical crossover "none with high integration", "agnostic with high integration" and "atheist with intermediate or high integration" are too small to be included in the table. This makes sense, as it is unlikely that a person calling him/herself an atheist or agnostic would consider him/herself highly integrated into a religion. Moreover, since the responses "rarely" and "never" were grouped together in the category "weak" integration, this also explains why the boxes for atheists, agnostics and people without religion but with weak or intermediate integration are not completely empty). Whether the individual is Catholic, Protestant, without religion, agnostic or atheist, the level of religious integration is related to or her level of scientific knowledge. The more integrated an individual, the lower his or her knowledge score.

**Table 9**
**Mean score for scientific knowledge by religion and level of religious integration (N=2037)**

| | |
|---|---|
| **Complete sample** | **4,49** |
| Strong Catholic Integration | 4,39 |
| Intermediate Catholic Integration | 3,98 |
| Weak Catholic Integration | 4,52 |
| Strong Protestant Integration | 4,00 |
| Intermediate Protestant Integration | 4,10 |
| Weak Protestant Integration | 4,54 |
| No Integration Intermediate | 3,78 |
| No Integration Weak | 4,67 |
| Agnostic Intermediate Integration | 6,04 |
| Agnostic Weak Integration | 6,29 |
| Atheist Weak Intégration | 7,16 |

The Pew Research Center survey asks a question specifically about the Evangelical movement that is absent from the European data and thus offers another interesting measure: "Would you describe yourself as an Evangelical Christian or a "born-again" Christian?" It is designed to determine how closely aligned an individual is to the evangelical movements in the United States that are characterized by an emphasis on the Holy Spirit and its spiritual dimension. 48% of respondents define themselves in this way. The majority of them are Protestant (86%) and 12% are Catholic. Belonging is also related to the variables of practice, prayer or the importance of religion in life. Thus, Evangelicals attend more church services than non-evangelicals (46% versus 21% go every week (Cramer's V= 0,27)), and pray more (71% versus 45% do so every day (Cramer's V= 0,27)). Consequently, 72% of them consider religion to be very important in their life (versus 32% for non-evangelicals (Cramer's V= 0,40)).

Being Evangelical is therefore likely to reinforce the influence of the other religious variables. This hypothesis is confirmed by Tables 10 and 11, which show that membership in an Evangelical movement, regardless of the degree of practice and the frequency of prayer, is associated with less scientific knowledge.

**Table 10**
**Average score for scientific knowledge according to religious practice and evangelical affiliation (N=1490)**

| Complete Sample | **4,09** |
|---|---|
| Weekly Practice, Non-Evangelical | 4,49 |
| Weekly Practice, Evangelical | 3,96 |

**Table 11**
**Average score for scientific knowledge according to prayer and evangelical affiliation (N=1338)**

| Complete sample | **4,17** |
|---|---|
| Daily Prayer, Non-Evangelical | 4,31 |
| Daily Prayer, Evangelical | 3,76 |

The importance accorded religion, a measure of the subjective dimension of religion, and membership in an Evangelical group is also related the level of scientific knowledge (Table 12). As with religious practice and prayer, Evangelicals, regardless of the importance accorded religion, perform less well than non-Evangelicals.

**Table 12**
**Mean score for scientific knowledge according to the importance given to religion and Evangelical affiliation (N=1483))**

| Complete Sample | **4,08** |
|---|---|
| Religion less important, not Evangelical | 4,40 |
| Religion less important, Evangelical | 3,96 |

As we observed in the European data, all of these results for Americans confirm the existence of a negative correlations between declared religions and level of religious practice on the one hand and the level of scientific knowledge on the other.

**2. The effect of age, level of education and gender on the level of scientific knowledge**

Now let's see how socio-demographic variables affect this conclusion. The control variables included in the surveys used in this paper were: age, gender (female and male) and level of education. Unfortunately, the data do not include level of revenue. This is a limitation but the effect of revenue is indirectly included through the level of education, which is well known to be strongly correlated with it. We will first analyze the European data (section 2.1) and then the American data (section 2.2).

    **2.1 The European Union**

The analysis of the European Union data can be summarized thus:

- Overall, men scored higher on the knowledge scale than women (8.77 versus 7.81). This result is independent of age and education level;

- Age does not discriminate. Up to the age of 55, the scores obtained for scientific knowledge are the same, whereas those aged 55 and over obtain lower scores. The strong correlation between the respondent's age and the age of graduation explains this result. The older the respondent is, the lower his or her school-leaving age. It is therefore the level of education rather than age that explains this relation;

- As expected, education is more significant than age or gender; as it increases, so does the average scientific literacy score (Table 13).

**Table 13**
**Average score of correct answers to the question on scientific knowledge by age of graduation**

|  | Mean | N |
|---|---|---|
| **Complete Sample** | **8,16** | **25008** |
| 16 years or less | 6,69 | 8397 |
| 17-19 years | 8,32 | 9135 |
| 20 years and over | 8,49 | 7476 |

The score differences between the least and most educated individuals are very large (6.69 versus 8.49). These results are not surprising insofar as the cognitive dimension of a societal phenomenon is very often related to the time spent in school. However, this variable does not eliminate the correlation between religion and the level of scientific knowledge (Table 14).

**Table 14**
**Mean Score by Religion and Age of Graduation**

| Religion | Level of Education | Mean | Number |
|---|---|---|---|
| Catholic | 16 years or less | 6,48 | 4538 |
|  | 17-19 years | 8,32 | 4548 |
|  | 20 years or more | 9,34 | 2670 |
| Orthodox | 16 years or less | 4,96 | 1171 |
|  | 17-19 years | 6,95 | 1264 |
|  | 20 years or more | 8,40 | 887 |
| Protestant | 16 years or less | 7,89 | 1126 |
|  | 17-19 years | 8,73 | 889 |
|  | 20 years or more | 9,92 | 1609 |
| Muslim | 16 years or less | 3,69 | 119 |
|  | 17-19 years | 6,59 | 88 |
|  | 20 years or more | 8,15 | 47 |
| No religion | 16 years or less | 8,44 | 871 |
|  | 17-19 years | 9,22 | 1616 |
|  | 20 years or more | 10,38 | 1591 |

Those without religion, even the least educated (having left school before the age of 16) have a higher score of scientific knowledge than the most educated Muslims (having left school at 20 or older).

Table 15, measure the "added value" of a change in the level of education on the level of knowledge for each declared religion. We observe that the effect of a change in level of education on the knowledge score is higher for Muslims, and much less lower for Protestants and non-religious people. This interesting result could be related to the extent to which each religion deny or limit the interpretive autonomy of individuals in matter of religious precepts and scientific theories. In all cases however, it remains that a higher level of education lead to a higher score of knowledge, though the effect of each education scale is different for each religion.

**Table 15**
**Added value of the level of education on the level of knowledge according to religion (N=23034)**

| | |
|---|---|
| Muslim | 4,46 |
| Orthodox | 3,44 |
| Catholic | 2,86 |
| Protestant | 2,03 |
| No Religion | 1,94 |

We did a logistic regression on the Eurobarometer data taking the scientific knowledge score as a dependent variable and the age of graduation and the respondent's religion as independent variables. We chose to test a model where "the non-religious" with a "school-leaving age above 19 years" are the reference group. The choice of this group is explained by the fact that it is composed of individuals most likely to have a high score of scientific knowledge. It should be noted that choosing another benchmark does not change the results.

A complementary regression analysis using "Muslims who left school at age 16 or younger" as the reference group confirms these results.

The model is significant (Pr > F less than .0001) and confirms the relevance of religion and education as explanatory variables. This simple model explains 18.9% of all observed variations. A regression analysis, based on the seven variables taken into consideration, shows that those without religion, with a school-leaving age over 19 years, obtain a score of 9.74. Declaring belonging to a religion and having "less education" lowered the average score. The decrease is negligible for Protestants and four times greater for Catholics. The maximum effect is observed for Muslims and Orthodox (Table 16). In sum, the Eurobarometer surveys indicate that among the main religions, only Protestant membership does not significantly affect the score of scientific knowledge.

**Table 16**
**Logistic regression analysis by religion and school-leaving age**

|  | Parameter values | Pr > |t| |
|---|---|---|
| **Reference group: no religion, education 19 years or older** | 9.74757 | <.0001 |
| Catholic | -1.24116 | <.0001 |
| Orthodox | -2.33504 | <.0001 |
| Protestant | -0.31448 | <.0001 |
| Muslim | -3.19462 | <.0001 |
| Education 16 years or less | -2.46421 | <.0001 |

The religion to which one belongs is therefore, all other things being equal, more statistically explanatory of the score of scientific knowledge, as measured by the questionnaire, than the age of graduation. Although the level of education has a strong influence on the results, it does not cancel out the specific effect of belonging to a religion.

The addition of other variables into the regression analysis does not change the previous results. Thus, taking gender (female or male), age and intensity of religious practice into account induces only minor variations (Table 17).

**Table 17**
**Logistic regression by religion, age, school leaving age, gender and religious practice**

|  | Parameter Values | Pr > \|t\| |
|---|---|---|
| **Reference group: no religion, education 19 years old or older, female, under 25 years old, non-practicing** | 9.49562 | <.0001 |
| Muslim | -3.27178 | <.0001 |
| Catholic | -1.08484 | <.0001 |
| Orthodox | -2.19970 | <.0001 |
| Protestant | -0.04295 | *0.5160* |
| Education 16 years or less | -2.06093 | <.0001 |
| Male | 0.77996 | <.0001 |
| 25-39 years | 0.09158 | *0.1144* |
| 40-54 ans | -0.10925 | *0.0590* |
| 55 ans et plus | -0.83543 | <.0001 |
| Pratiquant | 0.00836 | *0.8635* |

This model explains 22.2% of the variation, which is little more than the previous model. In other words, these three variables provide little added information for the explanation of the observed variation in the score of scientific knowledge and therefore do not significantly alter the specific influence of the religious variable.

It should also be noted that religious practitioners and the 25-39 and 40-54 age groups are not statistically significant, which is indicative of their lack of explanatory relevance. It should also be noted that Protestantism no longer appears as a relevant explanatory variable in this model. That religion, which was the least determinant of knowledge in the previous model, is no longer determinant in this model, which includes more variables. Subject to the "competition" of other variables, the specificity of Protestantism fades away; this seem consistent with what we know of the the particular characteristics of a Protestantism which, at the doctrinal level, allows for a great freedom of individual thought and was never systematically opposed to science and the autonomy of the scientific sphere with regards to the religious sphere.

In summary, the regression analyses confirm the previous results; religion and education level are important explanatory variables of scientific knowledge and they overshadow the other socio-demographic variables. While the level of education is positively correlated with higher knowledge scores, a strong religious identity is, in contrast, postively correlated with lower knowledge scores.

### 2.2 The United States

For the United States, Table 18 shows that, as could be expected from the above results, the level of education modulates the effect of religious affiliation on knowledge scores. Since the age of graduation is strongly positively correlated with the scientific knowledge score, the following hypothesis can be advanced: within the same religious group, respondents with a high level of education will have more knowledge than those with a low level of education. Regardless of the group considered, this hypothesis holds true. Indeed, when the respondent has a low level of education, the score is between 1.30 and 2.18 times lower than if he or she has a bachelor's degree or higher.

**Table 18**
**Mean score for scientific knowledge by education level controlled by religious belonging (N=2262)**

|  |  |  |
|---|---|---|
| Protestant or Catholic | **Entire Sample** | **4,00** |
|  | Bachelor's degree or higher | 5,49 |
|  | Post-Secondary Education | 4,16 |
|  | High school or less | 3,00 |
| No religion | **Entire Sample** | **4,21** |
|  | Bachelor's degree or higher | 6,06 |
|  | Post-Secondary Education | 4,86 |
|  | High school or less | 2,78 |
| Agnostic or Atheist | **Entire Sample** | **6,63** |
|  | Bachelor's degree or higher | 7,23 |
|  | Post-Secondary Education | 6,60 |
|  | High school or less | 5,58 |

A complementary hypothesis can be formulated: if the level of education is more predictive of the scientific knowledge score than the religious group to which one belongs, then agnostic or atheist individuals with a low level of education should have a lower score than Catholics or Protestants who have a high level of education. The results show that this hypothesis does not hold true, and it can therefore be concluded that belonging to a religious group is more explanatory of the knowledge score measured than the level of education.

The addition of the level of religious integration in the analysis slightly changes the intensity of the results, but does not change the direction of the relationship (Table 19).

From these results, it is possible to conclude that:

- The addition of the level of integration does not change the direction of the relationship: atheists or agnostics with a low level of integration and a low level of education perform better than highly integrated Catholics or Protestants with a high level of education;

- Regardless of the group considered, within the group, those with the lowest level of education have also the lowest score of scientific knowledge (the amplitude of the ratio of Bachelor's degree or higher / Secondary or lower varies between 1.25 and 2.44).

**Table 19**
**Mean score for scientific knowledge by level of education controlled by the index of religious affiliation and integration (N=2037)**

| | | | |
|---|---|---|---|
| Group 1 | Catholic or Protestant with strong religious integration | **Whole sample** | **4,08** |
| | | Bachelor's degree or higher | 5,55 |
| | | Post-secondary education | 4,05 |
| | | Secondary or less | 3,02 |
| Group 2 | Catholic or Protestant with intermediate religious integration | **Whole sample** | **4,06** |
| | | Bachelor's degree or higher | 5,55 |
| | | Post-secondary education | 4,17 |
| | | Secondary or less | 3,13 |
| Group 3 | Catholic or Protestant with weak religious integration | **Whole sample** | **4,53** |
| | | Bachelor's degree or higher | 6,05 |
| | | Post-secondary education | 4,57 |
| | | Secondary or less | 2,48 |
| Group 4 | No religion with intermediate religious integration | **Whole sample** | **3,78** |
| | | Bachelor's degree or higher | 5,67 |
| | | Post-secondary education | 4,40 |
| | | Secondary or less | 2,67 |
| Group 5 | No religion with weak integration | **Whole sample** | **4,67** |
| | | Bachelor's degree or higher | 6,39 |
| | | Post-secondary education | 5,43 |
| | | Secondary or less | 2,83 |
| Group 6 | Agnostic or atheist with intermediate religious integration | **Whole sample** | **6,17** |
| | | Bachelor's degree or higher | 7,03 |
| | | Post-secondary education | 6,20 |
| | | Secondary or less | 4,49 |
| Group 7 | Agnostic or atheist with weak religious integration | **Whole sample** | **6,76** |
| | | Bachelor's degree or higher | 7,27 |
| | | Post-secondary education | 6,72 |
| | | Secondary or less | 5,83 |

Table 20 shows that these graduation ratios are highest for groups 3, 4 and 5, i.e. the groups with the least religious integration. In other words, the stronger the religious integration, the smaller the effect of degree level on knowledge level.

**Table 20**
**Bachelor's degree or higher / Secondary or less for the seven "religious" groups (N=2037)**

| | |
|---|---|
| Group 1 : Catholic or protestant, strong integration | 1,84 |
| Group 2 : Catholic or protestant, intermediate integration | 1,77 |
| Group 3 : Catholic or protestant, weak integration | 2,44 |
| Group 4 : No religion, intermediate integration | 2,12 |
| Group 5 : No religion, weak integration | 2,26 |
| Group 6 : Agnostic or atheist, intermediate integration | 1,57 |
| Group 7 : Agnostic or atheist, weak integration | 1,25 |

Respondents who declare themselves to be Catholic or Protestant, but with a low level of religious practice and prayer activity (group 3) are more "culturally religious" than "religious by conviction". We can therefore conclude that the variable "religion" is not central to their lives.

In this case, as for people without religion, the level of education is more explanatory of scientific knowledge than the religious variable. For those who report having no religion, but who have an intermediate level of integration (group 4), the level of education is less explanatory of their score of scientific knowledge than for respondents in group 5. It can therefore be concluded that educational level is less predictive of the scientific literacy score than religious group affiliation.

## 3. Socio-demographic and religious determinants of representations of science

Having highlighted the links between religious beliefs and the level of scientific knowledge, the next section will analyses the relations between these beliefs and attitudes (positive or negative), towards science, first in European countries (section 3.1) and then in the United States (section 3.2).

### 3.1 Representations of science in the European Union

In order to identify the representations of science present among the people of the European Union, the Eurobarometer asks a series of questions on the social impact of scientific developments and opinions on science, technology and the environment. The questionnaire also includes a question on the degree of scientificity of various disciplines. We have not analyzed it here, as it is not very informative with regards to the relationship between science and religion. Respondents were asked to rank 13 professions or organizations to determine those which were best qualified to explain the impact of scientific developments on society (Table 21). Perceptions of course vary according to socio-demographic variables but here we focus on the religious variable.

**Table 21**
**Persons or organizations best qualified to explain the impact of science and technology on society (Percentage of mentions)**

| Complete Sample | 18 |
|---|---|
| University scientists | 53 |
| Television journalists | 34 |
| Scientists in the private sector | 28 |
| Print journalists | 27 |
| Physicians | 23 |
| Environmental organizations | 20 |
| Consumer groups | 12 |
| Writers and intellectuals | 10 |
| The government | 6 |
| Industry | 6 |
| Politicians | 5 |
| Representatives of religious groups | 2 |
| The army | 2 |

As shown in Table 22, the religion to which one belongs has an effect on the individuals or organizations considered most qualified to explain social impact of scientific or technological developments:

- Among the less educated, Orthodox individuals value university scientists and physicians, while Protestants and no religion tend to rely on consumer groups;

- Orthodox and no religion, at an intermediate level of education, value university scientists, while Protestants and Muslims value print journalists;

- Perhaps the most surprising variations are those affecting respondents with the highest school-leaving age. The no religion groups clearly favor – as do Orthodox and Muslims – university scientists. It can also be deduced from this table that, for all religions, schooling plays a key role in characterizing credible professions to explain the impact of scientific and technological developments on society.

In short, religion has an influence on the individuals or organizations deemed most competent to explain the impact of scientific and technological developments on society. Admittedly, the age of school leaving also affects opinions, but it never cancels out the effect of religion.

**Table 22**
**Individuals or organizations most qualified the impact of science and technology on society by education and religion of respondents (Percentage)**

|  |  | Television journalists | University scientists | Print journalists | Physicians | Scientists in the private sector | Environmental organizations | Consumer groups |
|---|---|---|---|---|---|---|---|---|
| 16 years or less | **Mean** | **40** | **39** | **26** | **24** | **22** | **15** | **11** |
|  | Catholic | 40 | 38 | 26 | 23 | 22 | 16 | 9 |
|  | Orthodox | 42 | **47** | 21 | **28** | 22 | 12 | 5 |
|  | Protestant | 40 | 32 | 27 | 26 | 19 | 15 | **17** |
|  | Muslim | 38 | 23 | 21 | 13 | 7 | 8 | 7 |
|  | No religion | 35 | 42 | 27 | 25 | 24 | 18 | **19** |
| 17-19 years | **Mean** | **35** | **55** | **27** | **24** | **30** | **20** | **11** |
|  | Catholic | 36 | 55 | 26 | 23 | 30 | 21 | 11 |
|  | Orthodox | 39 | **60** | 29 | 24 | **35** | 16 | 6 |
|  | Protestant | 37 | 44 | **32** | 27 | 26 | 20 | 14 |
|  | Muslim | **42** | 31 | **38** | 18 | 11 | 9 | 8 |
|  | No religion | 27 | **61** | 25 | 23 | 30 | 21 | 14 |
| 20 years or more | **Mean** | **30** | **62** | **31** | **22** | **32** | **23** | **16** |
|  | Catholic | 29 | 61 | 31 | 22 | 33 | 25 | 14 |
|  | Orthodox | 30 | **71** | 26 | **27** | **43** | 21 | 6 |
|  | Protestant | 34 | 53 | 33 | 23 | 26 | 20 | **21** |
|  | Muslim | 34 | **72** | 26 | 21 | 36 | 15 | 11 |
|  | No religion | 26 | **68** | 31 | 20 | 31 | 26 | 19 |

### 3.2 Representations of science in the United States

In the survey conducted in the United States, the questions concerning the representations of science are very different from those asked in the European Union analyzed above. It is therefore not possible to compare the results directly. However, as our focus is less on representations as such than on the possible relations between them and religious beliefs and practices, these differences do not create methodological problems. Thus, the correlations observed can be analyzed in the same way in both cases, even though the dimensions they measure may be different.

Seven questions, representing 11 variables, delineate the scope of our analysis. The first provides an overall view of the contribution (positive or negative) of science to society. 90% of the respondents believe that science has made life easier for most people, while only 10% believe the opposite. It should be noted in passing that, as the questions asked to assess the level of knowledge of Americans are different from those asked Europeans by the Eurobarometer, the fact that the relationship observed between the level of knowledge and religious beliefs is more or less the same in the two surveys strongly suggest that the two indicators of the level of knowledge do indeed assess the same underlying reality.

While age has little effect on results, women (88%) are less optimistic than men (93%). This difference between men and women remains, regardless of the level of education. The level of education is fairly strongly correlated with the overall perception of science. The higher the level of education, the more positive the perception of science is. For example, while 85% of the less educated believe that science has made life easier for most people, 96% of those with a bachelor's degree or higher agree with that statement.

Similarly, 91% of Americans surveyed said that science has had a mostly positive effect on health. This quasi-consensus is less clear-cut for the effect on the environment (76%) and on food (71%). For these two items, there is a critical minority who doubt the prevailing positivism about the benefits of science.

Unsurprisingly, responses to these three variables are closely correlated. A synthetic index of the positive effects of science shows that the average response is 2.36 out of a maximum of 3. We observe that 59% of the respondents think that science always has positive effects and only 6% think that it has none. Opinion is therefore overwhelmingly positive.

These answers are highly correlated with the previous question about whether or not science has made life easier. For example, 62% of respondents who say that science has made life easier consider that science has had three positive effects while only 19% of those who say that science has made life more difficult think so. Note that men are significantly more likely than women to attribute positive consequences to science (Table 23).

**Table 23**
**Number of positive elements according to socio-demographic and religious criteria**
**(Horizontal percentages)**

|  | Number of positive elements | | | |
|---|---|---|---|---|
|  | 0 | 1 | 2 | 3 |
| **Complete sample** | **6 %** | **12 %** | **24 %** | **59 %** |
| **Sex** | | | | |
| Male | 5 | 10 | 23 | **62** |
| Female | 6 | 13 | 25 | 55 |
| **Age** | | | | |
| 18-29 | 5 | 14 | 33 | 48 |
| 30-49 | 6 | 10 | 27 | 57 |
| 50-64 | 7 | 13 | 19 | 61 |
| 65 and over | 4 | 9 | 17 | **70** |
| **Diploma** | | | | |
| High school or less | 9 | 12 | 23 | 56 |
| Post-secondary education | 4 | 12 | 29 | 54 |
| Bacherlor's degree or more | 2 | 10 | 21 | **67** |
| **Religion** | | | | |
| Catholic | 5 | 10 | 23 | 62 |
| Protestant | 5 | 11 | 22 | 62 |
| No religion | 9 | 16 | 27 | 48 |
| Agnostic | 1 | 6 | 24 | **69** |
| Atheist | 2 | 6 | 16 | **76** |
| **Religious integration** | | | | |
| Weak | 3 | 10 | 22 | **65** |
| Intermediate | 7 | 12 | 24 | 57 |
| Strong | 5 | 12 | 27 | 56 |
| **Evangelist** | | | | |
| Yes | 7 | 11 | 24 | 58 |
| No | 4 | 11 | 20 | **66** |

Moreover, as age increases, the probability of being positive about the consequences of science increases. While only 48% of 18-29 years-old have three positive elements, 70% of those 65 and older have three. This likely reflects the greater sensitivity of youth to environmental issues. Similarly, but with less amplitude, those with higher levels of education are also more positive. Religious affiliation also plays a role: atheists and agnostics have a much more positive view of the effects of science than do Catholics and Protestants. Similarly, the level of religious integration and whether or not one is evangelical significantly affects attitudes towards science.

Views regarding the use of animals for scientific research divided respondents into two groups. While 47% support it, 53% oppose it (Table 24). Those who say that science has made life easier are more supportive of the use of animals (49% versus 35%). Similarly, those who think science has had three positive effects are more supportive of animal use than those who say science has had no positive effects (54% versus 25%). Furthermore, the ethical choice of whether or not to use animals varies according to socio-demographic and religious criteria.

While age does not appear to be a discriminating factor, men are more favorable to the use of animals for research purposes than women (59% versus 37%), as are the most highly educated (58% versus 41%). Finally, atheists stand out in clearly approving more than all other groups the use of animals (58%).

As was the case for the level of scientific knowledge, representations of science are affected not only by the individual's socio-demographic but also by their religious identity.

**Table 24**
**Favorable or unfavorable to the use of animals for research based on socio-demographic and religious variables (Horizontal percentages)**

|  | Favorable | Opposed |
|---|---|---|
| **Total** | **47** | **53** |
| **Sex** | | |
| Male | **59** | 41 |
| Female | 37 | **63** |
| **Age** | | |
| 18-29 | 45 | 55 |
| 30-49 | 51 | 49 |
| 50-64 | 45 | 55 |
| 65 and over | 48 | 52 |
| **Diploma** | | |
| Bachelor's degree or higher | **58** | 42 |
| Postsecondary education | 46 | 54 |
| High school or less | 41 | **59** |
| **Religion** | | |
| Catholic | 49 | 51 |
| Protestant | 50 | 50 |
| No religion | 37 | **63** |
| Agnostic | 47 | 53 |
| Atheist | **68** | 32 |
| **Religious integration** | | |
| Strong | 51 | 49 |
| Intermediate | 43 | 57 |
| Weak | 54 | 46 |
| **Evangelist** | | |
| Yes | 52 | 48 |
| No | 48 | 52 |

**Conclusion**

This secondary analysis of previous American and European surveys has shown that, everything being equal, there exists strong correlations between knowledge and representation of science on one hand and religion identity and practices on the other. For although the level of education obviously affects that relation it does not, however, cancel out the specific influence of the religion that is declared and practiced. One limitation of our regression analyses is that it does include the level of economic revenue and thus cannot measure the specific effect of that variable on the score of scientific knowledge. However, given that revenue is strongly correlated with the level of education, we do not expect that this variable would, if added, cancel the effect of religion, as revenue is already implicitly present in the education variable. But that must remain a hypothesis that could be tested with new data sets.

The major conclusion is thus that the "no religion" and Protestants tend to score better on the standard tests of scientific knowledge than Catholics, Orthodox or Muslims and that the result depend not only on the religious identity of respondents but also on the intensity of their religious practices. This latter association suggests that the more a religion exercises a strong imprint on its adepts, the less scientifically literate they tend to be. Representations of science also vary according to socio-demographic and religious variables and the latter are still negatively correlated with having positive attitudes towards science. Thus, the least religious Europeans and Americans also have the most positive representations of science and its social impacts.

Now, it is well known that correlation studies cannot identify causal links. However, it is also obvious that in the socialization process of individuals within families, religious education comes many years before scientific education. That temporal succession strongly suggests the most likely direction of the causal link. At the individual level, however, particular life experiences may later reverse the causality and scientific training may in turn feedback on religious conceptions and lead to abandoning some religious beliefs learned in a younger age that could then be seen as competing or in contradiction with the scientific knowledge acquired.

Most papers concluding that there is no necessary or even real conflict between science and religion are based on questions asking explicitly if there are or should be such conflicts. But the very asking of such questions may in fact bias the situation, given the tendency of most people to try to avoid conflict, and also the tendency to mix factual and normative questions. In contrast to such approaches, our analysis does not suffer from these limitations at it finds the links between religion and knowledge and attitudes towards science indirectly through statistical analysis of independent variables. It should also be noted that, to our knowledge, no study based on large-scale, empirical data, such as those analyzed here, highlights the opposite relation, namely that, at the macro-sociological level, religious beliefs and practices are positively correlated with what can be called the scientific attitude, here measured through knowledge scores and representations of the effects of science in society. Finally, it bears repeating that the existence of such macro-social realities does not exclude the obvious fact that some scientists can in their conscience positively associates science and their personal religious beliefs and practices and even promotes a "dialogue" between science and religion [13].